# Generating Fano Resonances in a single-waveguide nanobeam cavity for efficient electro-optical modulation

*Jianhao Zhang, [†,§] Xavier Le Roux, [†] Elena Ruran-Valdeiglesias, [†] Carlos Alonso-Ramos, [†] Delphine Marris-Morini, [†] Laurent Vivien, [†] Sailing He, [*,§] and Eric Cassan[*,†]*

[†] Centre for Nanoscience and Nanotechnology (C2N), CNRS, University Paris-Sud, University Paris-Saclay, 91405 Orsay cedex, France

[§]Centre for Optical and Electromagnetic Research, Zijingang Campus, Zhejiang University, Hangzhou 310058, China

ABSTRACT: We propose a method for generating Fano resonance in a standalone silicon nanobeam cavity which eliminates the inconvenience from the unexpected side-coupled bus waveguide and unlocks new opportunities to develop ultra-compact and ultra-fast electro-optical modulators.. Taking advantage from a spatial-division multiplexing principle of operation between transverse electric modes, a sharp resonant mode and an efficient flat background mode are simultaneously generated in the same silicon channel for the realization of efficient Fano resonances. Unambiguous asymmetric spectrum lineshapes are thoroughly investigated using numerical and analytical methods and experimentally demonstrated in the near infra-red around $\lambda$=1.55µm, presenting an extinction ratio of about 17dB for a $\Delta\lambda$ =56 pm wavelength detuning for the 1st cavity mode ($Q$-factor $Q$=34000), and higher than 23 dB extinction ratio for a $\Delta\lambda$ =366 pm detuning for the 2nd cavity mode ($Q$=5600). These extinction ratios are 10~15 dB larger than their Lorentzian counterparts exhibiting similar $Q$ factors. Silicon Fano modulation based on plasma dispersion effect is proposed, for which an energy consumption as low as few fJ/bit is estimated. Fano cavity scheme addressed in this paper presents a great potential for low power consumption silicon optical modulators and provides anew insight to the advantages of Fano resonances for optical modulation schemes.

Over the past decades, silicon has been determined to be the most promising platform for on-chip optical interconnects and nanophotonics. The technology for passive devices on the silicon-on-insulator (SOI) platform, such as beam splitters, polarization rotators, resonators, gratings for wavelength division multiplexing (WDM), and active devices like Si/Germanium based devices [1-3] have matured substantially. As key components, silicon optical modulators based on plasma dispersion effect have demonstrated high performances in terms of bandwidth, speed and extinction ratio[4]. However, the rapidly increasing demand for optical interconnects leads to the requirement of ultralow power consumption and tens-to-hundreds-gigabits devices.

Micro-resonators including micro rings/disks and photonic crystal cavities, which feature small footprint and relatively small driven signal for high extinction ratio, have been found to be a promising solution to enable on-chip silicon modulation[5-10]. and compact integrated photonic switches[11-13]. However, to ease the burden from the considerable photon lifetime in a high-quality factor (Q) cavity, resonators designed for high bit rate modulators have limited $Q$ factors of a few thousands, so that the photon decay rate can surpass the modulated bandwidth. This sacrifice on $Q$ factors results in a larger power consumption of tens to hundreds of femtojoules[14,15] for acceptable extinction ratio. Methods like different-signaled driving[16] and vertical P-N structures[17] for silicon disk resonators have been implemented to reduce to the power consumption. However, these kinds of doping schemes are complicated and challenging with respect to fabrication accuracy control. Therefore, new and simple solutions are expected for low power consumption and high-rate modulation.

In this context, Fano resonances, which arise from the interference of a discrete resonant mode and a continuum background[18], present efficient transitions between waveguide reflection and transmission compared to Lorentzian kind cavities. This extraordinary behavior can be used to address the bandwidth-power trade-off of silicon resonant modulators and potentially minimize the power consumption of silicon switching and modulation devices. Different types of Fano-resonance-based cavities have already been proposed including spatial membrane structures[19], plasmonic resonators[20] and integrated side-coupled one/two-dimensional (1/2D) photonic crystal cavities[21-24]. Thanks to the advances of fabrication technology, novel integrated devices based on these classical Fano cavities like asymmetric transmission[25], Fano lasers[26] and switches[27] have been demonstrated. Especially, an all-optical high-bitrate modulation behavior combining the free carrier response of indium phosphide and Fano cavity has been demonstrated[27,28]. Among earlier demonstrated integrated Fano cavities, most of previous structures consist of a bus waveguide and a side coupled photonic crystal cavity[21-24]. Considering the width-sensitive coupling and the structure-induced inconvenience of such side-coupled nanobeam resonators, on-chip electro-optical modulation relying on the control of free carrier concentrations in doped structures reveals to be challenging and has not been demonstrated yet. The flexible use of P-N junctions[29] was demonstrated in 2D photonic crystal cavities but the high insertion loss of such configurations overshadows their advantages. Fano resonance was also observed in a single nanobeam cavity[30] but Fano generation mode mixing between interfering channels was only made at the collection fiber level stage, degradation of Fano generation was observed with the fiber location, and design methodology for flexible and systematic Fano generation process still remain unsolved.

In this context, we address here the investigation of the combination of nanoresonators and control electrodes to develop a new class of Fano resonance silicon modulators aiming at a drastic reduction in power consumption without sacrificing bandwidth. Design, fabrication and characterization of sharp and effective Fano resonances behavior in a simple standalone waveguide structures for optical switching and modulation is reported. First, we present the mechanism used for realizing an effective Fano resonance in a single waveguide geometry without any side-coupled bus waveguide that is classically necessary in Fano cavities[20-28]. Then, we systematically investigate the behavior and robustness of Fano spectra by combining numerical and analytical methods and experimental characterizations. Our analysis stems from a quantitative explanation , gives new insight to the advantages of Fano resonances  for optical modulation schemes, and indicates room for further optimization.Combining rib nanobeam waveguide structures[31] or nano-arms-assisted nanobeam resonators[32], the reported new Fano cavity scheme can be expanded for the development of low power consumption optical modulators based on the plasma dispersion effect.

**PRINCIPLE**

A classical configuration of integrated Fano cavity is shown in Fig. 1 (a). Basically, it consists of a bus waveguide and a side-coupled micro/nano cavity providing both the continuum and discrete resonances needed to the Fano interference mechanism[21]. A partially transmitting element is placed in the waveguide to provide a control of the amplitude of the continuum path. The input mode is separated into both discrete levels and continuum physical channels, recombining and interfering at the output port, respectively.

The principle to obtain a Fano resonance is, basically based on the capability to separate the incident mode into two paths that can interfere at the output, one and the other being characterized by narrow ("discrete level") and wide ("continuum") spectral responses, respectively. We thus propose here a solution consisting in using a two-mode spatial multiplexing principle, taking care of the control of the spectral responses for each of both modes and to force their interference at the output. In this configuration, the two modes share the same physical channel, i.e a single optical waveguide. This concept is shown in Fig. 1 (b). The standalone Fano cavity consists of a single-mode input waveguide, a nanobeam cavity which cross section supports two transverse electric-field (TE) modes, and a subwavelength mode mixer. The TE1/TE2 modes, which are schematically depicted by the blue/red dash curves in bottom left inset of Fig. 1 (b), have symmetric/asymmetric fields within the waveguide cross-section, respectively. The Multimode Interference (MMI)-like operation allows a tight control of the balance between the excited TE1 and TE2 modes at the narrow/wide waveguide transition through the choice of the narrow (input) and wide (nanobeam) waveguides' widths, labelled $w_i$ and $w_n$, respectively (see Fig. 1 (b)). The key of the design is also to fix the effective indices of the two modes at sufficiently different values so that the TE1 mode strongly feels the influence of the cavity, thus generating a marked spectral resonance, while the second mode TE2 does not see practically the patterned holes of the waveguide and thus presents a very flat transmission spectrum (i.e. an ultra-wide resonance). The spectrum of both TE1 and TE2 modes are presented by the solid curves in Fig. 1 (b) at position 3. The nanobeam cavity is followed by a subwavelength mixer for inter mixing between the TE1-mode narrow resonance and the TE2-mode wide resonance, respectively, which produces two Fano resonances, i.e. for each to both TE modes. An extra narrow side waveguide is then placed close to the nanobeam cavity waveguide to form a directional coupler for extracting the TE2 mode and its conversion into the TE1 mode of the side waveguide (see the left inset of Fig. 1 (b)). Five planes marked by black dash lines (1,2, 3, 4, and 5) at different positions in Fig. 1 (b) are used to illustrate the structure principle of operation: plane 1 at the input waveguide, plane 2 before the cavity, plane 3 before the mode mixer, plane 4 after the mode mixer, and plane 5 at the bifurcation of the output directional coupler.

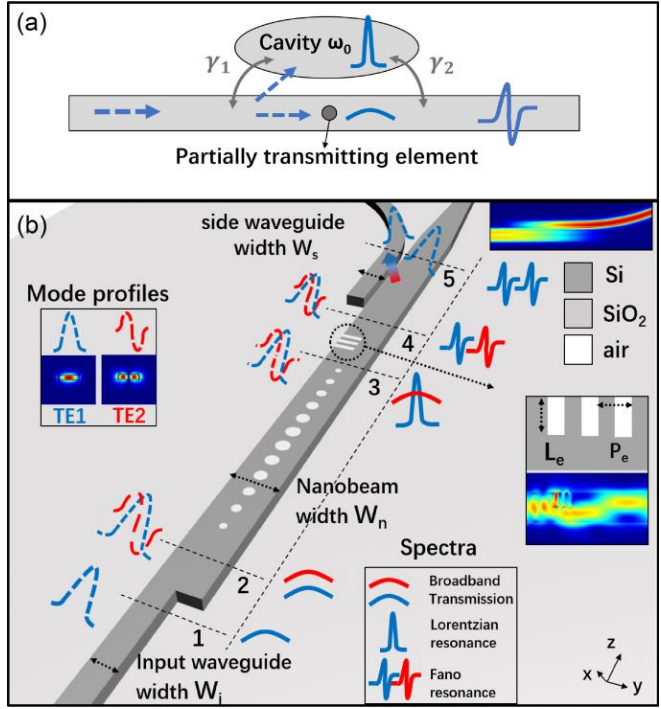

**Figure 1.** (a) Schematic of classical integrated Fano cavity. $\gamma_1$ and $\gamma_2$ are the decay rates due to cavity modes-/waveguide modes coupling. (b) Schematic of the proposed standalone Fano cavity, consisting of a MMI-like input structure, a nanobeam cavity, a subwavelength mixer and a directional coupler. The blue and red dashed curve represent the spatial mode profiles (bottom right inset) of the TE1 and TE2 mode, respectively. The blue and red solid curves represent the spectral lineshapes of the TE1 and TE2 modes, respectively. Top right inset: the propagating distribution of the TE2 mode coupled with and converted into the TE1 mode of the side waveguide. Middle right inset: zoom-in schematic of the 3-slit subwavelength mixer and the propagating distribution corresponding to a TE1 mode injection.

A configuration corresponding to a typical silicon-on-insulator (SOI) platform with a 220nm thick silicon core and 2μm thick buried silicon dioxide layer was considered to apply these concepts. The dispersion curves of both TE1 and TE2 modes (at plane 2) obtained for different nanobeam widths ($w_n$) are presented in Fig. 2 (a). To provide a high transmission level for the TE2 mode with a weak perturbation from the nanobeam cavity, the difference of effective index values between the TE1 andTE2 modes has to be as large as possible to prevent any Bragg reflection for the TE2 mode. A waveguide width $w_n$=630 nm ($n_{eff,TE1}$~2.55 and $n_{eff,TE2}$~1.6) corresponds to the ideal value but due to the rapid change of the TE2 mode dispersion close to this condition, this option can lead to fabrication-sensitivity issues. Thus, a moderate value $w_n$=800nm was selected, providing an acceptable index contrast between the TE1 and TE2 modes of ~ 0.565. Next, the nanobeam width $w_n$ being fixed at 800nm (throughout this work), the excitation efficiency of both TE modes in the nanobeam waveguide cross-section was adjusted by varying the input waveguide width $w_i$. As shown in Fig. 2 (b), the excitation efficiency of the TE1 mode, i.e. the energy of the TE1 mode in the nanobeam waveguide normalized to that in the input waveguide (at plane 1) increases with the waveguide width, while the excitation efficiency of the TE2 mode first reaches a maximum at $w_i$=400nm, and then gradually decreases with increasing $w_i$. These trends are due to the increasing effective index of the guided mode and mode mismatches between of the input and the nanobeam waveguide modes. However, according to Fig. 2 (a), the effective index matching for both TE1 and TE2 modes occurs for $w_i$ close to 400nm for TE1 and 800nm for TE2. In addition, the $w_i$=400nm case provides the largest mode asymmetry at the input/nanobeam waveguide interface, which favors the generation of the asymmetric mode (TE2 mode). Fig. 2 (b) indicates that the excitation efficiency of the TE1/TE2 mode can be adjusted in a large range (from nearly 55%/40% to infinity), which is convenient for the control of the Fano line shape by balancing the TE1/TE2 ratio (as shown hereafter). For input waveguide widths larger than 400nm, the summation of the TE1 and TE2 excitation efficiencies (i. e. the summation of blue and red curves in Fig. 2 (b)) is close to 1, which also indicates a proper width range for low-loss injection.

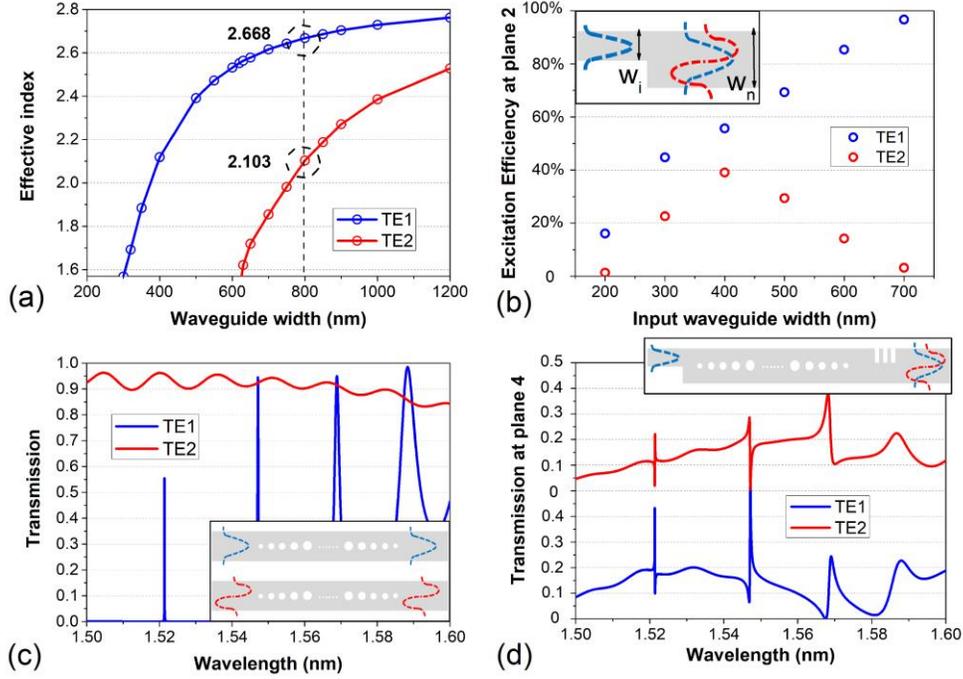

**Figure 2.** (a) Dispersion curves of the TE modes at 1550nm wavelength, in a 220nm-thick silicon on insulator strip waveguide with different width $w_n$ (plane 2 in Fig. 1 (b)). (b) Excitation efficiencies of the TE1 and TE2 modes in a nanobeam waveguide with $w_n$=800 nm, connected with an input waveguide with different widths $w_i$ ranging from 200-700nm. (c) Transmission spectra (from plane 2 to plane 3 in Fig. 1 (b)) of the TE1 and TE2 modes through a classical nanobeam cavity with $w_n$=800nm, 50 holes and 300nm period. (d) Transmission spectra of the TE1 and TE2 modes through the proposed Fano cavity with a 400nm-wide input waveguide. The period, filling factor and length of the subwavelength holes array are $P_e$=200 nm, $ff$=0.5 and $L_e$=400nm, respectively. Other parameters are identical to that reported in (c). In each figure, the TE1 and TE2 modes are depicted by blue and red curves solid lines or circles, respectively. Structures corresponding to the (b), (c) and (d) configurations are shown in the insets, in which the TE1 and TE2 spatial mode profiles are displayed in blue and red dash lines, respectively.

For the nanobeam cavity, a 50-holes array under a period of 300nm was considered. To obtain a resonant high quality ($Q$) factor and a high transmission for the TE1 mode, the hole radii were tapered from 100nm in the center to 70nm in a 15-periods length[33]. Extra 10-holes mirror sections with identical 70nm radius holes were added at the end of each taper. The total length of the nanobeam cavity was lower than 15 μm. The quality factor and the transmission were calculated using 3-dimension Finite-Difference-Time-Domain (3D FDTD) simulations. The calculated transmission spectra (from plane 2 to plane 3 in Fig. 1 (b)) for both TE1 and TE2 modes are shown in Fig. 2 (c). Since the hole radius profile near the cavity center slightly varies, the studied cavity presents 4 resonant modes in the 1500-1600 nm wavelength range. The simulated quality factor/transmission values for the 1$^{st}$ and 2$^{nd}$ cavity modes are 230000/50%, and 19000/80%, respectively. Considering that high photonic lifetimes are generally not expected in modulation structures, the estimated $Q$ values is already large enough. In contrast, the nanobeam cavity provides high and broadband transmission for the TE2 mode and an average transmission larger than 90% is obtained in the wavelength range from 1500nm to 1600nm. Though a balanced transmission of these two modes is needed for perfect interferences that the response of the nanobeam cavity does not fully satisfy (the maximum transmission of the 1$^{st}$ cavity mode of TE1 mode is about 40 % smaller than the average transmission of TE2 mode), the related excitation ratio can be easily engineered by adjusting the input waveguide width of the MMI-like structure to find the right balance between the two optical waves.

To mix both TE1 and TE2 modes after the nanobeam cavity (plane 4 in Fig. 2(b)), which is probably the most essential point for Fano mode generation, a subwavelength mixer consisting of few rectangles (3, only, are enough in the typical studied configuration) made of asymmetrically-located etch holes (Fig. 1 (b)) is placed after the nanobeam cavity. The period, filling factor, and length of the hole are $P_e$=200 nm, $ff$=0.5 and $L_e$=400nm, respectively. Both TE modes in the nanobeam waveguide excite the first two order modes of the subwavelength structure, which are converted back into the TE1 mode again after the mixer. By carefully selecting the mixer length, we achieve a TE1-TE1 and the TE2-TE1 mixing efficiency of ~45% and ~35%, respectively. These coupling efficiencies are obtained here from a 3-N (3 holes) subwavelength mixer (optimizing methodology for the subwavelength mixer is presented in supplementary materials). The simulated propagating distribution corresponding to TE1 injection is shown in the middle right inset in Fig. 1 (b). The mode interference behavior after the subwavelength mixer is an obvious signature of multimode generation. The transmissions of TE modes through the complete device (inset in Fig. 2 (d)) including the input waveguide width $w_i$=400nm, the MMI-like structure (Fig. 2 (b)), the nanobeam cavity (Fig. 2 (c)), and the subwavelength mixer, are reported in fig. 2(d).

Unambiguous Fano line spectra for the TE1 mode (blue curve in Fig. 2 (d)) and the TE2 mode (red curve in Fig. 2 (d)) are observed, which confirms the proper interference mechanism of the resonant and the flat spectra. Interestingly and as it could be anticipated, the TE2 mode (red curve in Fig. 2 (d)) also exhibits a Fano spectrum lineshape, since part of the TE1 and TE2 modes are indeed coupled to the TE2 mode as well after the mixer. Uncomplete destructive interferences marked by the appreciable but not complete Fano dips (transmission does not drop to 0) result from the unbalanced TE1/TE2 energy levels. Each of both Fano resonance behaviors can be separately optimized by adjusting either the TE1-TE1, TE2-TE1 mixing efficiencies or the TE2-TE2, TE1-TE2 ones.

**ANALYTICAL DESCRIPTION OF FANO RESONANCE**

To further study the generation of Fano resonances in such standalone cavity structures, we performed complementary analytical calculation using the temporal coupled-mode theory. The equivalent schematic view of the cavity dynamics is shown in Fig. 3. The total structure can be considered as a two-port scattering system. $S_I^+$, $S_I^-$ are the forward and backward field amplitudes from port I (left side port), respectively. $S_{II}^+$ and $S_{II}^-$ share the same definitions for port II (right side port). $T_{F1} = |t_{F1}|^2$ is the total transmission of the TE1 mode after the mixer (energy of TE1 mode in plane 4 normalized to that at plane 1). For a nanobeam cavity with a resonant frequency $\omega_1$ and corresponding electric field $a_1$ for the TE1 mode, the decay rate for this resonant mode due to coupling to the two feeding waveguides, the decay rate due to vertical out coupling and intrinsic absorption are $\gamma_1, \gamma_2, \gamma_v, \gamma_i$, respectively. Therefore, the total decay rate for TE1 mode resonance can be written as $\gamma_t = \gamma_1 + \gamma_2 + \gamma_v + \gamma_i$. The energy excitation efficiency from TE1 mode to the TE1 and TE2 modes in the MMI-like structure (defined in Fig. 2 (b)) are labeled by $\eta_1$ and $\eta_2$, respectively. Meanwhile the efficiencies from the TE1 and TE2 modes to the TE1 mode of the subwavelength mixer at the output port are labelled as $C_{11}$ and $C_{12}$, respectvely. The transmission of the nanobeam cavity for the TE2 mode called $T_2$ ($T_2 = |t_2|^2$) can be assumed as uniform due to the weak interaction of this mode with the array of patterned holes.

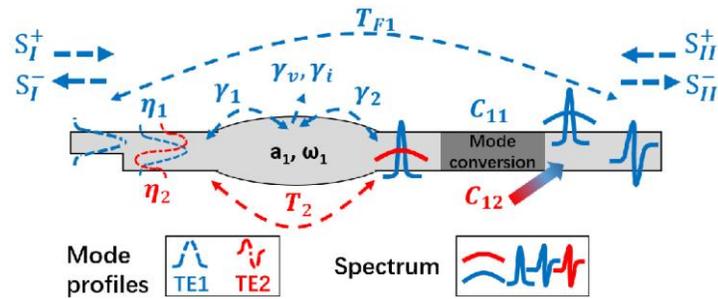

**Figure 3.** Schematic of the dynamics of optical waves in a standalone two-port waveguide nanobeam cavity. TE1 and TE2 mode is marked by blue and red dashed curves, respectively. The spectrum of TE1 and TE2 mode are depicted by the blue and red curves, respectively.

According to the temporal coupled-mode theory[21] and considering a frequency detuning of $\delta_r = \omega - \omega_0$, we established the coupled equation of the resonant and backward as follows:

$$\frac{da(t)}{dt} = (-i\delta_r - \gamma_t)a + \sqrt{2\gamma_1}e^{i\theta_1}\sqrt{\eta_1}S_I^+ \tag{1}$$

$$S_I^- = r_m\sqrt{\eta_1}S_I^+ + \sqrt{2\gamma_1}e^{i\theta_1}a \tag{2}$$

$$S_{II}^- = \sqrt{2\gamma_2}e^{i\theta_2}a\sqrt{C_{11}} + \sqrt{\eta_2}S_I^+ t_2\sqrt{C_{12}}ie^{i\Delta\theta} \tag{3}$$

In which $\theta_1$ and $\theta_2$ are the phase factors of the cavity waveguide modes in the left and right ports, respectively. $\Delta\theta$ is the phase difference between the TE1 and TE2 modes, which is counted from the MMI-like interface to the front plane of the mixer. The reflectance of nanobeam mirrors for the TE1 mode is $r_m$ (for a high-Q cavity, $|r_m|$ can be considered as 1).

For steady state condition, $da(t)/dt=0$ and the field transmission coefficient for the TE1 mode through the whole structure ($T_{F1} = |t_{F1}|^2$) can be written as:

$$t_{F1} = \frac{S_{II}^-}{S_I^+} = \frac{2\sqrt{\gamma_1\gamma_2\eta_1 C_{11}}e^{i(\theta_1+\theta_2)}}{i\delta_r+\gamma_t} + i\sqrt{\eta_2 C_{12}}t_2 e^{i\Delta\theta} \tag{4}$$

For symmetric cavity design, $\theta_1 = \theta_2$, $\gamma_1 = \gamma_2$, and for a highly confined cavity mode, $e^{i(\theta_1+\theta_2)} = \cos(\theta_1+\theta_2) + i\sin(\theta_1+\theta_2) = -r_m \approx -1,$[26] therefore

$$t_{F1} = \frac{S_{II}^-}{S_I^+} \approx \frac{-2\gamma_1\sqrt{\eta_1 c_{11}}}{i\delta_r+\gamma_t} + i\sqrt{\eta_2 C_{21}}t_2 e^{i\Delta\theta} \tag{5}$$

If there is no excitation of the TE2 mode and no mixer region after the cavity ($\eta_1 = c_{11} = 1$, $\eta_2 = C_{12} = 0$), then

$$T_F = |t_t|^2 \approx \frac{4\gamma_1^2}{\delta_r^2 + \gamma_t^2} \tag{6}$$

The total transmission of the TE1 mode then goes back to a typical Lorentzian shape.

For a case where $\eta_2 \neq 0, C_{21} \neq 0$:

$$t_{F1} \approx \frac{\frac{-2\gamma_1}{\gamma_t}\sqrt{\eta_1 C_{11}}}{i\frac{\delta_r}{\gamma_t}+1} + i\sqrt{\eta_2 C_{12}}t_2 e^{i\Delta\theta} = \frac{\frac{2\gamma_1}{\gamma_t}\sqrt{\eta_1 C_{11}}\left(i\frac{\delta_r}{\gamma_t}-1\right)}{\frac{\delta_r^2}{\gamma_t^2}+1} + i\sqrt{\eta_2 C_{12}}t_2 e^{i\Delta\theta} \tag{7}$$

$$T_{F1} \approx \left(\frac{\frac{-2\gamma_1}{\gamma_t}\sqrt{\eta_1 C_{11}}}{\frac{\delta_r^2}{\gamma_t^2}+1} - \sqrt{\eta_2 C_{12}}t_2 \sin\Delta\theta\right)^2 + \left(\frac{\frac{2\gamma_1}{\gamma_t}\sqrt{\eta_1 C_{11}}}{\frac{\delta_r^2}{\gamma_t^2}+1}\frac{\delta_r}{\gamma_t} + \sqrt{\eta_2 C_{12}}t_2 \cos\Delta\theta\right)^2 \tag{8}$$

$$= \frac{\frac{4\gamma_1^2}{\gamma_t^2}\eta_1 C_{11}}{\frac{\delta_r^2}{\gamma_t^2}+1} + \eta_2 t_2^2 C_{12} + \frac{\frac{4\gamma_1}{\gamma_t}t_2\sqrt{\eta_1 C_{11}\eta_2 C_{12}}}{\frac{\delta_r^2}{\gamma_t^2}+1}\left(\frac{\delta_r}{\gamma_t}\cos\Delta\theta + \sin\Delta\theta\right)$$

Assume that the phase relative variable is written as $C_p = -\frac{\delta_r}{\gamma_t} + \frac{\delta_r}{\gamma_t}\cos\Delta\theta + \sin\Delta\theta$, we simplify $T_t$ as:

$$T_{F1} \approx \eta_2 t_2^2 C_{12}\left(\frac{\frac{4\gamma_1^2\eta_1 C_{11}}{\gamma_t^2\eta_2 C_{12}t_2^2}}{\frac{\delta_r^2}{\gamma_t^2}+1} + \frac{\frac{\delta_r^2}{\gamma_t^2}+1}{\frac{\delta_r^2}{\gamma_t^2}+1} + \frac{\frac{4\gamma_1\sqrt{\eta_1 C_{11}}}{\gamma_t t_2\sqrt{\eta_2 C_{12}}}\left(\frac{\delta_r}{\gamma_t} + C_p\right)}{\frac{\delta_r^2}{\gamma_t^2}+1}\right)$$

$$= \eta_2 T_2 C_{12}\frac{\left(\frac{\delta_r}{\gamma_t} + \frac{2\gamma_1\sqrt{\eta_1 C_{11}}}{\gamma_t t_2\sqrt{\eta_2 C_{12}}}\right)^2}{\frac{\delta_r^2}{\gamma_t^2}+1} + \eta_2 T_2 C_{12}\frac{\left(1 + \frac{4\gamma_1\sqrt{\eta_1 C_{11}}}{\gamma_t t_2\sqrt{\eta_2 C_{12}}}C_p\right)}{\frac{\delta_r^2}{\gamma_t^2}+1} \tag{9}$$

Assuming that $\frac{\delta_r}{\gamma_t} = \epsilon$, $\frac{2\gamma_1\sqrt{\eta_1 C_{11}}}{\gamma_t t_2 \sqrt{\eta_2 C_{12}}} = q$, and taking into account that the total energy of the TE2 mode coupled back to TE1 mode is $T_{12} = \eta_2 T_2 c_{12}$, we can simplify A(10) as:

$$T_t \approx T_{12} \frac{(\epsilon+q)^2}{\epsilon^2+1} + T_{12} \frac{(1+2qC_p)}{\epsilon^2+1} \tag{10}$$

The first term of equation (10) is exactly close to the traditional Fano expression, which is used for describing a two-ports Fano cavity[34]. The variable $q$ is the asymmetric parameter, which quantifies the Fano spectrum asymmetry, $q=\pm 1$ corresponding to perfect Fano resonances in classical Fano cavities. The normalized energy amplitude of the continuum part of the TE1 mode, e.g. the $T_{21}$, becomes the amplitude coefficient of the Fano spectrum in Eq. (2). Similarly the classical Fano spectrum in which the amplitude coefficient is the transmission of the partially transmitting element[18]. However, the above equations indicate that the phase quantity also contributes to the transmission and causes a small deviation to the perfect Fano line shape. By controlling $C_p$, this phase related item can be minimized and high quality Fano lineshapes can then be obtained. Most importantly, the above analytical calculation shows that the obtained pseudo Fano expression opens room to design a sharp Fano-like behavior. Analysis for TE2 mode can be made similarly and is not shown.

To study the effect of excitation efficiencies and mixing efficiencies to the spectrum and for simplification, we assume that $\Delta\theta = 0$, then $C_p = \epsilon(-1 + cos\Delta\theta) + sin\Delta\theta = 0$, we can simplify (10) to:

$$T_t = T_{12} \frac{(\epsilon+q)^2}{\epsilon^2+1} + T_{12} \frac{1}{\epsilon^2+1} \tag{11}$$

According to equation (11), we then adjust the energy ratio of the TE1/TE2 modes and the mixing efficiency to find the optimization operation point. The details are shown in Figs. 4 (a) and (b). In Fig. 4 (a), the resonance wavelength, the transmission of the TE2 mode $T_2$ in nanobeam, the quality factors $Q_1$ due to coupling to feeding waveguide $\gamma_1$ and $Q_2$ due to the vertical coupling $\gamma_v$, and the intrinsic absorption $\gamma_i$ are 1550nm ($\omega_0 = 193.414 THz$), 90%, $Q_1 = Q_2 = 7 \times 10^4$, $Q_v = 1.6 \times 10^5$, respectively. We assume that there is no absorption in the cavity and according to $Q_1 = \frac{\omega_1}{2\gamma_1}$ and $Q_v = \frac{\omega_1}{2\gamma_v}$, we obtain a total quality factor of $Q = \frac{\omega_1}{2\gamma_t} \approx 30000$. Meanwhile, the mixer coupling efficiencies from the TE1 and TE2 modes to the TE1 mode are adjusted to $C_{11} = 0.45$, $C_{12} = 0.35$, respectively. We change the excitation efficiency of the TE1 mode $\eta_1$ from 10% to 99.9% considering that there is nearly no loss in the MMI-like structure and the excitation efficiency is $\eta_2 = 1 - \eta_2$. We then calculate the related transmission curves and asymmetric parameter $q$ which are shown in Fig. 4(a). In these plots, clear Fano curves presenting sharp spectral transitions and high extinction ratio are demonstrated from $\eta_1 = 0.1$ and $\eta_1 = 0.9$. The best configuration for Fano behavior, indicated by $q \approx 1$, is obtained for $\eta_1 = 0.55$, as shown in Fig. 3. Especially, the comparison between the $\eta_1 = 0.9$ and $\eta_1 = 0.999$ cases shows that values of $\eta_1$ close to 1 are necessary to recover a Lorentzian shape cavity spectral lineshape.

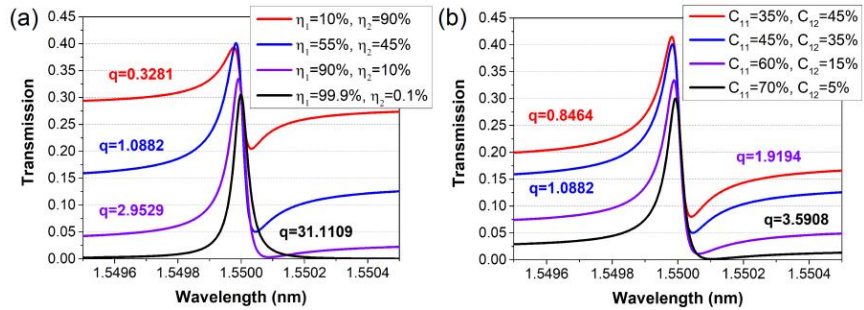

**Figure 4.** (a), (b) The transmission of the TE1 mode versus the varying excitation efficiencies $\eta_1$, $\eta_2$ and mixing efficiencies $C_{11}$, $C_{12}$.

We also study here the effect of the mixer with different coupling efficiencies ($C_{11}$ and $C_{12}$) on the Fano spectra, as shown in Fig. 4 (a). The parameters (quality factors, wavelength etc.) of the nanobeam are identical to those reported in Fig. 4 (a) while the excitation efficiency for TE1 mode is chosen as $\eta_1 = 0.55$. The coupling efficiencies $C_{11}$; $C_{12}$ of the mixer are arranged as 35%;45%, 45%;35%, 60%;15%, 70%;5%. Tthese efficiencies are obtained from 3-N, 5-N, 10-N and 15-N subwavelength mixers, which behaviors are detailed in supplementary materials. The *q* parameter variation is much more limited than in the previous case and the spectra keep a fairly marked Fano shape even in the severe $C_{11}$; $C_{12} = 70\%;5\%$ condition, which indicates the robustness of the proposed Fano cavity scheme against variations of the mixer geometry. Besides, comparing and summarizing the main trends reported in Fig. 4 (a) and (b), we can conclude that the peak transmission drops with the increasing TE1 efficiency, i.e. the excitation efficiency $\eta_1$ and the mixing coupling efficiency $C_{11}$, and at the same time with larger deviations of *q* from 1. This trend, to some extent, provides a design strategy to target a trade-off between correct Fano lineshapes and large extinction ratio values.

**EXPERIMENTS**

The scanning electron microscopy (SEM) picture of a typical fabricated device is shown in Fig. 5. The gap (Fig. 1 (b)) between the nanobeam waveguide and the side waveguide ($w_s$=400nm) is 150nm. A straight coupling length of 16 μm was chosen to completely couple the TE2 mode in the nanobeam waveguide the TE1 mode in the side waveguide. Then the side waveguide was turned into a bend waveguide with a radius of 40 µm for a good separation of both modes. The performances of this directional coupler were confirmed by 3D-FDTD simulations and from many fabricated devices characterizations. The mixer dimensions consisted of 3 rectangle etched holes with a 200nm period and a 50% filling factor, each rectangle hole having a length ($L_e$ in Fig. 1 (b)) of half $W_n$, i. e. of 200nm. The width of the input waveguide was chosen at 500nm since the experimental transmission of resonant mode was usually lower than the theoretical one. Other parameters (e. g. nanobeam waveguide with $W_n$, numbers of holes, holes radii and holes tapering scheme) about the nanobeam cavity were directly considered from the design stages.

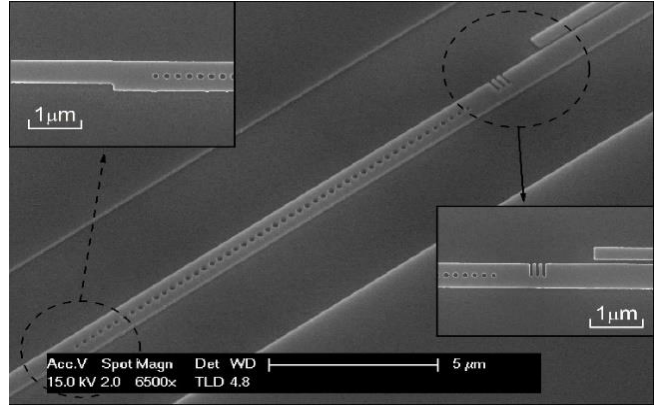

**Figure 5.** (a) SEM of Fabricated devices. The MMI-like structure and subwavelength mixer is shown in the top-left and bottom-right inset, respectively.

The transmission curves of the full structure are shown in Fig. 6 (a). Clear Fano lineshapes for the 1st and 2nd order cavity modes are observed, with resonance wavelengths located at around 1515 nm and 1531 nm, respectively, which turned to be coincident with the 3D simulation reported in Fig. 2 (d). Simulated distributions for these two modes are presented in the insets of Fig. 6 (a). Simultaneous Fano lineshapes of the 2nd cavity mode in the nanobeam and side waveguides are also shown in Fig. 6. (b), as the blue and orange curves, respectively. The details of 1st and 2nd cavity mode in the nanobeam waveguide output shown in Fig. 6 (a) are presented in Fig. 6 (c) and (d), respectively. The blue circles trace and the orange solid curves are experimental results and fitting curves from Eq. (10), respectively.

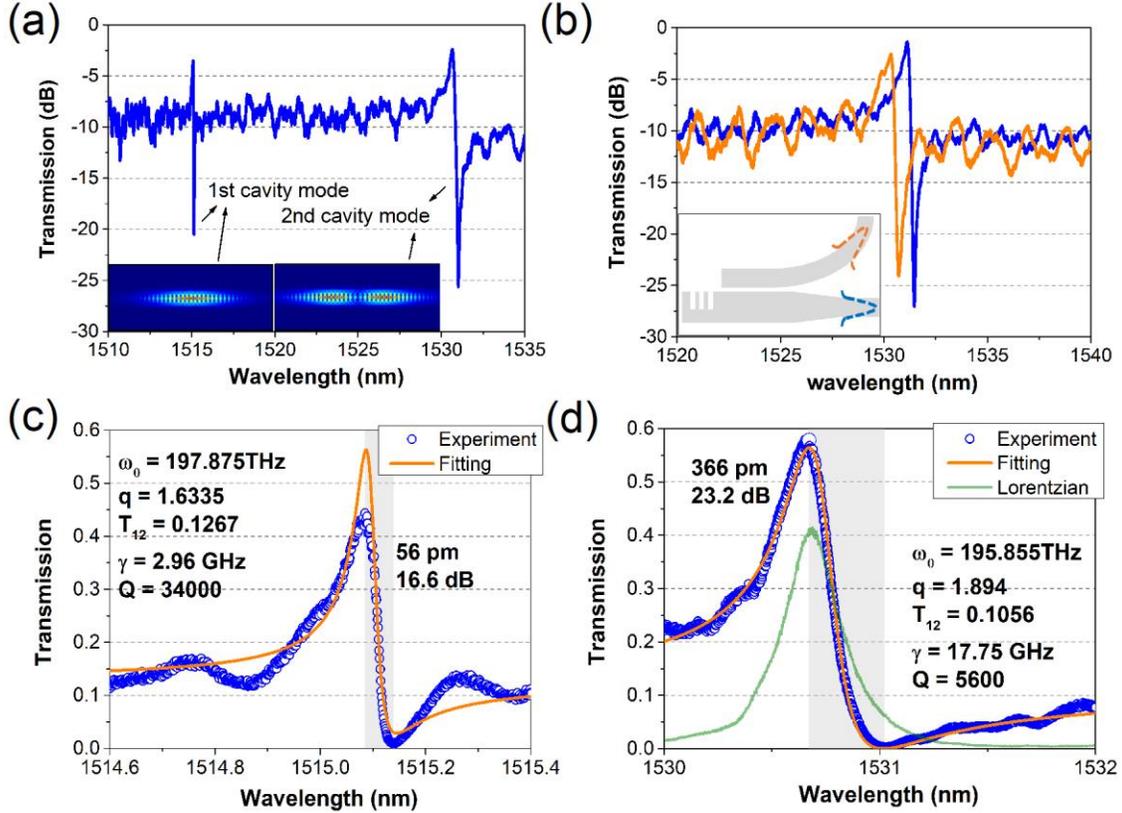

**Figure 6.** (a) Experimental transmission of the nanobeam waveguide. (b) Experimental transmission around 2nd cavity mode detected in nanobeam and side waveguides, depicted by blue and orange curves, respectively. (c) Experimental transmission and fitting curve of 1st cavity mode of nanobeam waveguide, depicted by blue circles and orange curve, respectively. (d) Experimental transmission and fitting curve of 2nd cavity mode of nanobeam waveguide, depicted by blue circles and orange curves, respectively. Lorentzian curve is labeled by a green curve. Transition between maximum and minimum are depicted by grey regions.

In Fig. 6 (c), we see that in a wavelength detuning of 56pm, the cavity optical transmission experiences a transition drop of about 17 dB. To experimentally obtain the quality factor, another device with same parameters but an input width of 700nm was further analyzed. This wide-input device can, be considered as a classical nanobeam cavity without any TE2 mode excitation (Fig. 2 (b)) in the MMI-like structure. According to our previous analysis, the transmission spectrum is then pretty close to a Lorentzian resonance. Using $Q = \frac{\Delta\omega}{\omega_0}$, a $Q$ factor of ~32000 was obtained (the transmission of this Lorentzian spectrum is low and not shown in Fig. 6 (c)). The determination of the $Q$ factor value was also extracted from Eq. (10), fitting the experimental Fano spectrum. The asymmetric total decay rate $\gamma_t$ was 2.96 GHz, which indicated a $Q$ factor of 34000, well coincident with that given by the Lorentzian-kind device. The asymmetric parameter $q$ was estimated as $q$=1.6335, i.e. close to the perfect Fano condition (i. e. $|q|\approx 1$) proving the consistency of the carried-out optimization. In addition, the total measured energy transmission from the TE2 mode to the TE1 one ($T_{12} = \eta_2 T_2 c_{12}$) was found around 0.1267. Considering the excitation ratio of TE2 mode ($\eta_2$) for a 500nm width input waveguide of about 0.3 and the nanobeam transmission ($T_2$) of 0.9 and a mixing efficiency ($c_{21}$) of 0.35, this value of $T_{21}$ is thus in good agreement with analytical prediction.

Similar analysis for the 2$^{nd}$ cavity mode was also performed and is presented in Fig. 6 (d). With a wavelength detuning of 366pm, more than 23.2 dB extinction ratio was obtained. The calculated asymmetric parameter $q$, the $Q$ factor and the total transmission $T_{12}$ were 1.894, 5600 and 0.1056, respectively (the phase variables $C_p$ for both cases were calculated to be close to zero and are not shown). An experimental Lorentzian spectrum with nearly the same $Q$-factor of 5600 is also reported (green curve) in Fig. 6 (d) for comparison. For the Fano spectrum, only 366 pm is needed for more than 23 dB extinction ratio. However such a high extinction ratio in a Lorentzian cavity requires up to a 2nm wavelength shift. We can notice here that a classical nanobeam cavity needs to raise its *Q* factor up to ~40000 to reach the same extinction efficiency. Considering the unexpected long lifetimes in micro cavities[14], the performance of our cavities exhibits an excellent potential for Fano resonance-induced efficient modulation and switching devices, i.e. far beyond classical Lorentzian nanobeam active cavities.

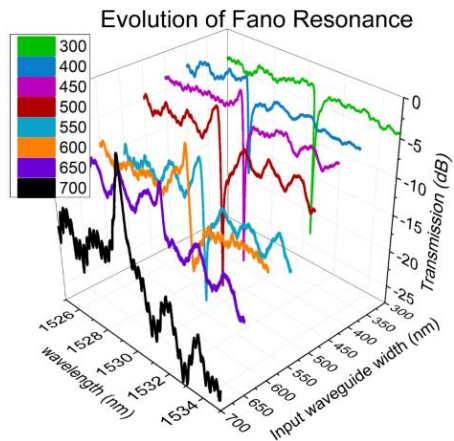

**Figure 7.** Evolution of the Fano spectrum by varying the width of the input waveguide $w_i$ (see Fig. 1 (b)) from 300nm to 700nm.

An array of Fano cavities with different input waveguide widths $w_i$ ranging from 300 nm to 700 nm was fabricated to measure the evolution of the Fano resonanceas a function of $w_i$. As shown in Fig. 7, the spectrum lineshapes first behave as all-pass filters and becomes asymmetric (trends to be Fano spectrum lineshape) with increasing input waveguide widths. The lineshape achieves almost a maximum asymmetry within the range of 500-600 nm of input waveguide width and comes back to an add-drop-type Lorentzian shape at $w_i \approx 700$ nm. Clear Fano lineshapes can thus be seen in a wide width range, i. e. from $w_i$=450nm to 650 nm. This easiness highlights the robustness of our design even in the case of possible tens of nanometers of fabrication errors.

**DISCUSSION**

Remembering that the limited quality factor (for high bit rate modulation) in a Lorentzian active cavity puts a practical ceiling to the extinction ratio and tends to increase the needed operating voltage swings (usually up to few volts[4]), the most important merit of Fano resonance resonators is to improve the extinction ratio and power consumption by providing a sharp transition between reflection and transmission states. Since our design is intrinsically suitable for P-N modulator and switching integrated schemes, one significant further work is here to estimate the power consumption of electro-optic modulation structures based on the plasma dispersion effect and relying on the proposed standalone waveguide Fano scheme.

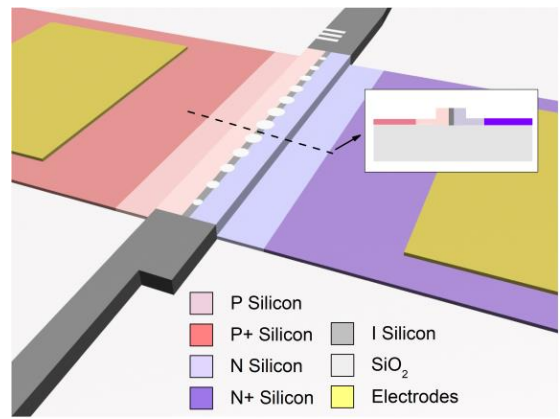

**Figure 8.** Proposed Fano modulator based on rib silicon P-N depletion structure. Inset is the cross-section of the rib structure at the position labeled by dot line.

The proposed P-N Fano modulator is shown in Fig. 8. and the analysis is performed on the 220nm SOI platform with 170 rib thickness and 50 nm slab thickness. The Fano cavity is slightly doped in both side and a depletion region is formed in the center. Meanwhile, the free carriers induced index change in silicon can be described at 1.55µm wavelength by[35]:

$$\Delta n = \Delta n_e + \Delta n_h = -[8.8 \times 10^{-22} \times \Delta N_e + 8.5 \times 10^{-18} \times (\Delta N_h)^{0.8} \quad (12)$$

$\Delta n_e, \Delta n_h, \Delta N_e$ and $\Delta N_h$ are here the electron induced index change, hole induced index change, the density changes of electrons, and density change of holes. For a common doping change level of $\Delta N_e = \Delta N_h = 5 \times 10^{17}$, the index change is $\Delta n = -1.7 \times 10^{-3}$. We consider a nanobeam cavity with symmetric P-N junctions, i. e. with identical widths for the P-type and N-type regions. The optical index in the P doped/depletion regions can be considered as 3.4783 and ~3.48, respectively. Since the depletion width of the P-N junction can be described as[36]:

$$W_d = \left[\frac{2\varepsilon_0\varepsilon_r(N_D+N_A)}{q(N_D \cdot N_A)}\left(V_P + V_{ext} - \frac{2k_BT}{q}\right)\right]^{1/2} \quad (13)$$

$\varepsilon_0 = 8.8542 \times 10^{-12} F/m$ and $\varepsilon_r = 12$, $V_P$, $V_{ext}$ and $q$ are the vacuum silicon dielectric constant, the silicon dielectric constant, the build-in potential, external applied voltage and the unit charge. $k_B$ is the Boltzmann constant and $T$ is the temperature fixed at 300K.

For a 0.35nm shift of the Fano resonance (consistent with the 2nd cavity mode discussed above, with $Q\sim5600$ and an extinction ratio larger than 20dB), the required width change of the depletion region ($\Delta W_d$) calculated by using 3D FDTD simulation was estimated to about 15$nm$. The peak-peak voltage for such a width change at a bias of $V_B \approx -0.5V$ is $\Delta V_{ext} \approx 0.5V$ (i. e. $V_{ext}$ is within the range of $V_B - \Delta V_{ext}/2$ to to $V_B + \Delta V_{ext}/2$). Considering an average depletion width of 100nm and a cavity length L=15$\mu m$, the capacitance of the P-N junction is[36]:

$$C_d = \varepsilon_0 \varepsilon_r \frac{H_{rib}(\sim 220nm)}{W_d} L \approx 4fF \qquad (14)$$

Therefore the Energy consumption per bit can be estimated to[14]:

$$E = \frac{C \cdot V_{pp}^2}{4} = \frac{C_d \cdot \Delta V^2}{4} \approx 0.25 fJ/bit \qquad (15)$$

Even considering a practical capacitance (that an experimental capacitance taking other capacitance and fabrication imperfection into account besides the depletion capacitance) of $C$=50fF like for recently reported ring resonators[14], the Energy consumption per bit is still as low as $3.125 fJ/bit$. Such a low energy consumption per bit is impossible to be achieved in resonant modulators with a similar low $Q$ of only 5600, let alone an extinction ratio as high as in this Fano cavity (23dB). For the 1st cavity mode of $Q$~34000, (0.056nm shift for ER>15dB), the energy consumption per bit is decreased down to $E \approx 0.5 fJ/bit$. Besides, asymmetric doping profiles could be further considered to optimize the active structure performances[37].

**CONCLUSION**

In this paper, Fano resonance is achieved in a standalone silicon nanobeam cavity without any side-coupled bus waveguide, which opens huge potential for the development of optical and optoelectronic devices including integrated modulators and switches. By smartly using the mode-diverse response of a single nanobeam cavity, controlling the spatial-division multiplexing and mode mixing in the multimode waveguide section, we realize a sharp resonant mode and a flat background mode in the same silicon channel for flexible generation of Fano resonances. Unambiguous asymmetric spectrum lineshapes were experimentally observed, presenting 16.6 dB extinction ratio within 56 pm wavelength detuning for the 1st cavity mode ($Q$- factor $Q$=34000) and 23.2 dB extinction ratio within 366 pm detuning for the 2nd cavity mode ($Q$=5600), which are much better than their Lorentzian counterparts with similar $Q$ factors. We explored the mechanisms of the Fano spectrum line shapes in single-waveguides. Our investigation based on an analytical model and numerical calculation gave clear and quantitative explanation, suitable for optimization of the pseudo Fano resonances. The energy consumption of active modulation or switching single waveguide Fano cavities using the plasma dispersion effect was estimated to be less than few fJ/bit, giving competitive potential to further low power consumption silicon optical modulators operating at high-data bit rates (>>1Gbits $s^{-1}$). We believe that the proposed Fano cavity scheme addresses the deadlock of integrated cavities for efficient optical modulation in single silicon waveguide geometries.

## METHODS

**Fabrication and Characterization.** A silicon-on-insulator (SOI) wafer with 220nm silicon thickness was used for fabrication. Device fabrication was based on Electro-Beam lithography, ICP-RIE etching and wet cleaning process[38]. A tunable laser (Yenista TUNICS – T100S) operating in the 1520-1640nm wavelength range, a polarizer, and a power meter (Yenista CT400), were used for charactering the devices. A couple of grating couplers are used for fiber-chip in and out interfacing. All the transmission curves are normalized to straight waveguides with identical grating couplers.

## ASSOCIATED CONTENT

**Supporting Information**

The Supporting Information is available free of charge on the ACS Publications website at DOI:

10.1021/acsphotonics. Xxxxxxx.

Optimization and robustness of subwavelength mixer.

## AUTHOR INFORMATION

**Corresponding Author**

*E-mail: eric.cassan@u-psud.fr

**Author Contributions**

J. Z. designed the devices, performed calculation and simulation, characterized the samples. X. L. R. and E. R. V. and C. A. R. fabricated the samples. D. M. M. and L. V. provided guidance in analyzing. J. Z. and E. C. wrote the manuscript with input from everyone. E. C. and S. H. supervised the work.

**Notes**

The authors declare no competing financial interest.

## ACKNOWLEDGMENT

We thank China Scholarship Council for supporting this work. The French ANR agency is also acknowledged for the its support through the SITQOM project.## REFERENCES

(1) Ravi, P.C. Academic and industry research progress in germanium nanodevicesIntegrated germanium optical interconnects on silicon substrates. *Nature* **2011**, 479, 324-328.<ск>26</ск>